\documentclass{ceab}   

\usepackage{epsfig}     
\usepackage{graphicx}   
\usepackage{hyperref}
\usepackage{siunitx}

\usepackage{natbib}     
\usepackage[T1]{fontenc} 
\def\runningtitle{Rodolfo Barb\'a Meeting}   
\def\articlenumber{7}
\def\ppages{1--34}

\begin{document}

\title{OB associations: from stellar to galactic scales}

\author{Alexis L. Quintana$^{1}$
\vspace{2mm}\\
\it $^1$Departamento de Física Aplicada, Facultad de Ciencias, Universidad de Alicante, \\ \it Carretera de  San Vicente s/n, 03690 San Vicente del Raspeig, Spain}

\maketitle

\begin{abstract}
Low-density and unbound stellar groups, OB associations have been historically delineated through their bright and massive members. They have been analysed for decades, but the arrival of Hipparcos, and more recently of \textit{Gaia} led to a change of paradigm by allowing the identification of more reliable members using parallaxes and proper motions. This renewed interest offers an opportunity to emphasize the role of OB associations across many areas of astronomy. In this review, I highlight their importance across multiple scales: how OB associations constitute suitable sites to study massive stars and stellar multiplicity, their relation with star clusters, their interactions with the interstellar medium through the feedback of their massive members, and how they shape the structure and evolution of the Milky Way and beyond.
\end{abstract}


\section{Context}

This review is adapted from the talk I presented at the meeting to honor Rodolfo Barbá. Unfortunately, I never had the chance to know him, as he passed away before I attended my first conference in-person. This meeting helped me to understand the impact he made in the astronomical community: Rodolfo had a plethora of research interests and managed to work with scientists from several fields. 

However modest may be my contribution, I am walking on the path traced by brilliant women and men over the past decades, including from Rodolfo Barbá. At the intersection of several fields in astronomy, OB associations notably connect the studies of massive stars, binarity, stellar clusters and Galactic structure; I thereby hope that, by highlighting the links between these fields, I can honor him in my own way.

\section{Introduction}
\label{intro}

\citet{Ambartsumian1947} defined OB associations as gravitationally unbound and low-density ($<$ 0.1 M$_{\odot}$ pc$^{-3}$) stellar groups, recognizable by their bright OB members. They have extensively been studied over the last $\sim$75 years, such that their main properties can be summarized as follows:

\begin{itemize}
    \item Because low-density groups are vulnerable against the tidal forces exerted from the Galactic potential \citep{Bok1934}, OB associations must be in the process of expanding \citep{Blaauw1964}. This expansion, whenever observed, is often anisotropic (e.g. \citealt{Wright2016,CantatGaudin2019}). 
    \item The fact that OB associations retain some kinematic coherence despite being unbound entails their youth. Decoupled from their birth environment \citep{LadaLada2003}, OB associations should be at least 1 Myr old, and no more than a few tens of Myrs \citep{Blaauw1964}, after which they will dissolve into the Galactic field population of stars \citep{Wright2020}.
    \item Typically extending from a few pc to a few tens of pc (e.g. \citealt{Blaauw1964,BlahaHumphreys1989,GarmanyStencel1992,MelnikEfremov1995,Gouliermis2018}), OB associations also carry a total mass of a few thousands M$_{\odot}$ to a few tens of thousands M$_{\odot}$ (e.g. \citealt{Wright2020,QuintanaWright2021,Quintana2023}), and their velocity dispersion is generally equal to a few km s$^{-1}$ \citep{WardKru2018,MelnikDambis2020,Quintana2023}.
    \item Assigning a single age to OB associations is a challenging task as they carry a significant level of substructure \citep{Ambartsumian1949,Blaauw1964,Ratzenbock}, composed of groups with various ages and kinematics (e.g. \citealt{Wright2014,PecautMamajek2016,Kounkel2018,CantatGaudin2019}).
\end{itemize}

\begin{figure}[h]
    \centering
    \includegraphics[scale =0.15]{ClusteredModel.pdf}
    \includegraphics[scale =0.2]{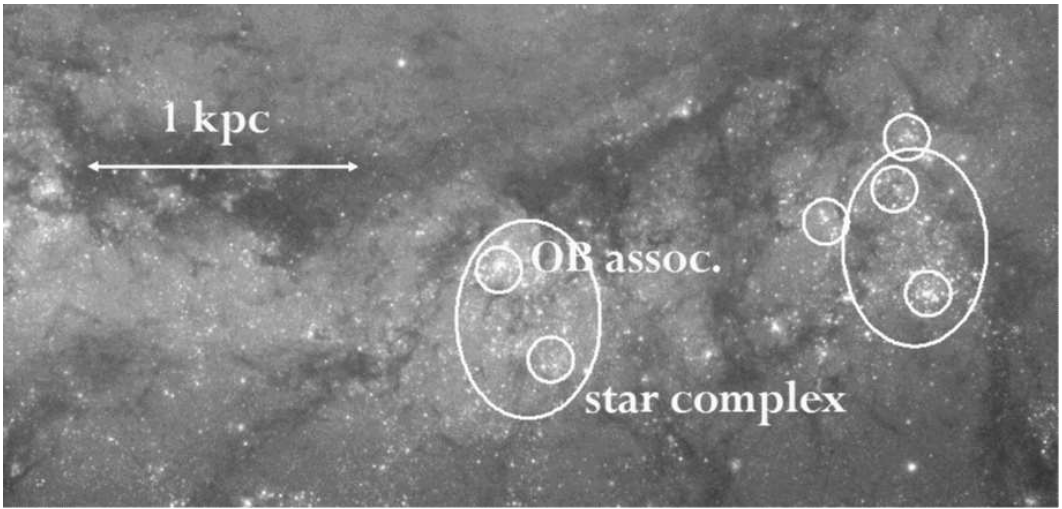}
    \caption{Top panel: Formation and disruption of a star cluster, briefly visible as an unbound association \citep{LadaLada2003} (Credit: Nick Wright). Bottom panel: Hierarchical structure within a star-forming complex (here within the inner spiral arm from M51) from \citet{Elmegreen2011b}, reproduced with permission from EDP Sciences. Note: the scales involved are different.}
    \label{OriginsOBAssoc}
\end{figure}

The origin of OB associations remains debated up to this day, with two models contrasting each other\footnote{There is also a third, more recent model, proposed by \citet{Krause2018}. In this scenario, called `surround and squash', recently-formed massive stars generate superbubbles which break out the parent elongated molecular cloud, before surrounding and squashing the denser parts of the gas, prompting more star formation.}. The \textit{clustered} model finds its roots in the review from \citet{LadaLada2003}, wherein the preponderance of stars (70--90\%) originate from dense, bound and \textit{embedded} clusters. As soon as massive stars form, they will disrupt the molecular gas used to form stars: consequently, only a minority of clusters (4--7\%) will survive as bound open clusters \citep{LadaLada1991}. The other clusters, unbounded by such feedback (a phenomenon referred to as \textit{residual gas expulsion}), will be briefly visible as low-density OB associations, before merging into the Galactic field. This model assumes that the embedded cluster is in virial equilibrium thanks to the combined gravitational potential of the gas and the stars. However, once the gas is taken away, the stellar part of the system will be found in a super-virial state and, consequently, expands \citep{Hills1980,Lada1984}. Conversely, the \textit{hierarchical} model envisions star-forming complexes as entities of various scales and densities, where open clusters arise from the densest regions, contrary to OB associations. The advantage of this model is that it does not require \textit{residual gas expulsion} to unbind the groups \citep{Heyer2001,Elmegreen2008,Kruijssen2012}. As often, the reality likely lies between the two models, but they are nonetheless useful to contrast. Since both scenarios are physically allowed, they will probably both occur in nature. They are illustrated in Figure \ref{OriginsOBAssoc}.

\section{Historical overview}

The field of OB associations has developed over the span of the last three quarters of century: a general outlook therefore appears timely. I have chosen to divide the history into three eras: \textit{Pre-Hipparcos}, \textit{Pre-Gaia}, \textit{Since Gaia}. I invite the readers interested in further details to visit the reviews from Adriaan Blaauw \citep{Blaauw1991}, Anthony Brown \citep{Brown1999} and Nick Wright \citep{Wright2020,Wright2023}, focusing on each era. I will personally delve more into the kinematic aspects of this area.

\subsection{Pre-Hipparcos}

Whilst OB associations were named for the first time in the mid 20th century, their history predates their label by a few decades. Low-density groups containing OB stars had been known since at least the beginning of the 20th century (e.g. \citealt{Kapteyn1914, Rasmuson1921}). Likewise, the instability of low-density stellar groups against the Galactic tidal forces was discovered more than a decade before the term \textit{OB associations} was used for the first time \citep{Bok1934}. 

This first era lasted for about half a century, during which several pioneers carried on the foundations laid by Viktor Ambartsumian \citep{Ambartsumian1947,Ambartsumian1949}. These researchers established the properties of OB associations (such as their age and spatial extent) and catalogued their members, with the most notable contributions from Adrian Blaauw \citep{Blaauw1956,Blaauw1964,Blaauw1991}, Jaroslav Ruprecht \citep{Ruprecht1966,Ruprecht1981}, Roberta Humphreys \citep{Humphreys1978,HumphreysMcElroy1984,BlahaHumphreys1989} and Catharine Garmany \citep{Garmany1973,GarmanyStencel1992,Garmany1994}. 

In these papers, overdensities of bright massive stars were identified on the plane-of-the-sky, entirely relying on photometric and spectroscopic observations of these objects. With the techniques available at the time, the membership of the classical lists of OB associations ended up heavily biased towards the earliest (O-B5) and visually brightest ($V < 5$ mag) stars. However essential were these works, they lacked the precise kinematic information required to define a robust membership for OB associations. 

\subsection{Pre-Gaia}

At the time of its release, the Hipparcos catalogue contained the most extensive ($>$ 100,000) list of measured stellar parallaxes \citep{ESA1997,Perryman1997}, and this led to an important development as parallaxes have been challenging to measure in the history of astronomy. Amidst other pivotal developments, this meant that OB associations could be characterized with the parallaxes and proper motions from this telescope.

\begin{figure}[h]
    \centering
    \includegraphics[scale = 0.25]{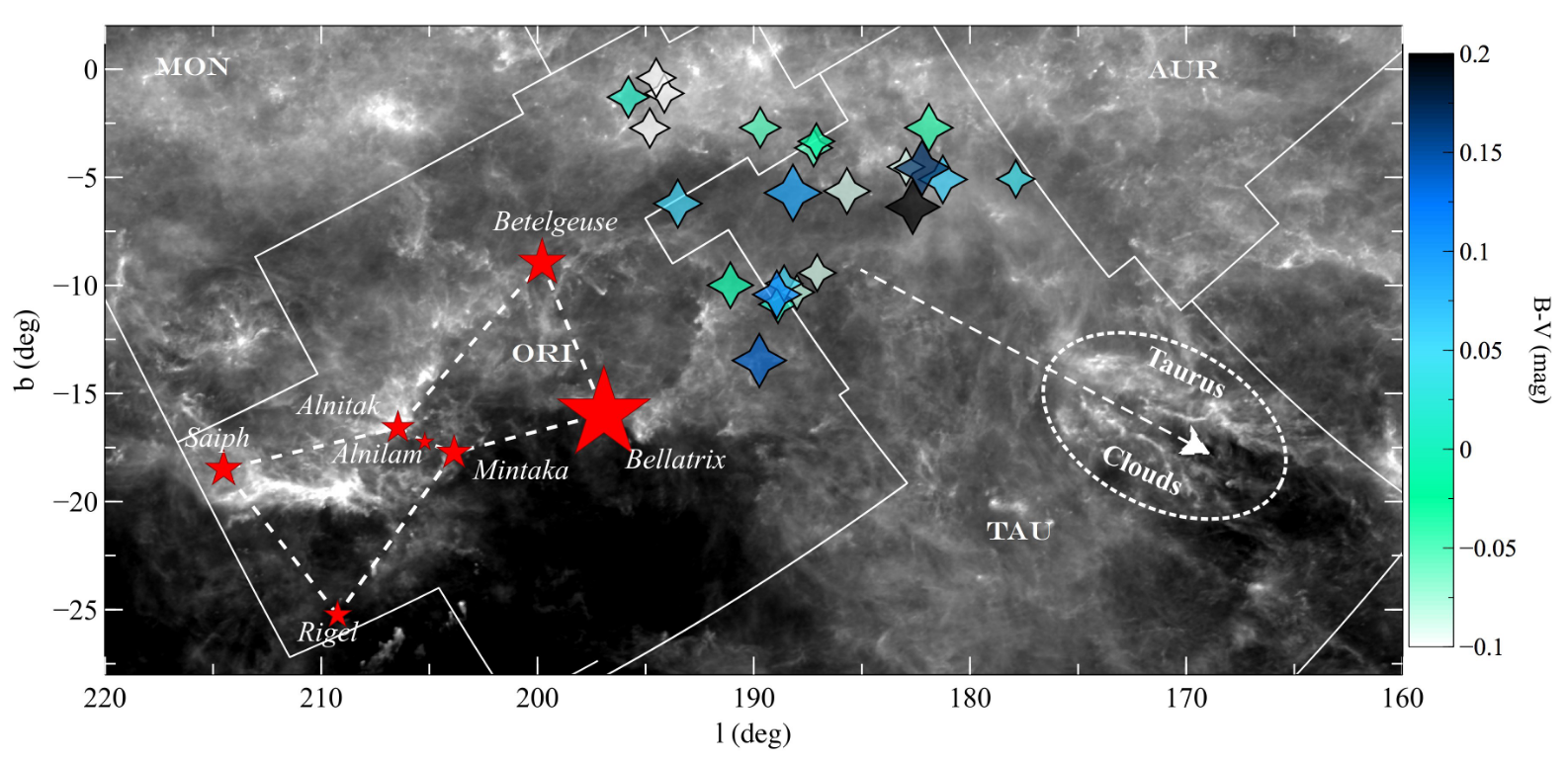}
    \caption{B-type members from Taurion colour-coded as a function of their B-V colour, in Galactic coordinates. The red symbols correspond to the giants and supergiants of the Orion constellation, while the background map displays the Taurus molecular cloud. This figure has been taken from \citet{BouyAlves2015} and reproduced with permission from EDP Sciences.}
    \label{Taurion}
\end{figure}

Hipparcos thus enabled a new census of nearby OB associations, among which Sco-Cen, Vela OB2, Trumpler 10, Cas-Tau, Cep OB2, and Per OB2 particularly benefited. Members as late as F-type could be included for the first time in an extensive way up to a distance of 650 pc (e.g. \citealt{HipparcosCensus1999,Elhias2006,MelnikDambis2009}). The structure and population of OB associations could be analysed in more details (e.g. \citealt{DeBruijne1999}) and appropriate techniques (e.g. the vector point method that exploited proper motions from Hipparcos) were applied to identify them and derive their properties such as their kinematic age \citep{Brown2000}. This second era ended with the identification of OB groups in the foreground of well-studied OB associations Orion OB1 and Vela OB2: these are Vela OB5, Monorion, Taurion and Orion-X in \citet{BouyAlves2015}, see Figure \ref{Taurion}.

In spite of this significant progress, the limits rapidly became apparent, as the available parallaxes in Hipparcos were mostly restricted to the solar neighborhood. Only a subsequent telescope could unlock an unbiased view of OB associations across the Galaxy, beyond the nearest ones. 

\subsection{Since Gaia}

The \textit{Gaia} mission \citep{Gaia} has provided the most comprehensive and accurate 3D map of the Milky Way at the time of writing. Its third data release includes $\sim$1.5 billion sources with astrometric data \citep{GaiaDR3}, four orders of magnitude greater than Hipparcos, truly revolutionizing the field of OB associations. The availability of such a wealth of data prompted new studies at an unprecedented scale, with groundbreaking methods and techniques developed over the recent past. 

Below are delineated the three main approaches that have been followed to study OB associations using \textit{Gaia} data:

\begin{enumerate}
    \item \textbf{Analysing the historical lists of OB associations with \textit{Gaia} data:} \citet{MelnikDambis2017} and \citet{MelnikDambis2020} investigated the kinematics in the historical catalogue of OB associations from \citet{BlahaHumphreys1989} with \textit{Gaia} DR1 and DR2, respectively. Only a handful of the existing OB associations were found to exhibit a significant expansion pattern, hinting at the lack of reliability of the historical membership of OB associations. 
    \item \textbf{Identifying compact subgroups within well-studied OB associations:} With its precise spatial and kinematic information, \textit{Gaia} is suited to identify substructures within OB associations, hence why this approach has been favoured, particularly in nearby associations such as Sco-Cen and Ori OB1 (e.g. \citealt{Kounkel2018,Squicciarini2021}). Expansion signatures tend to be more easily measured within kinematically-defined subgroups (e.g. \citealt{CantatGaudin2019,Armstrong2020}). Further details on this will be provided in Section \ref{subgroupsoc}.
    \item \textbf{Completely redefining OB associations:} Given the lack of kinematic coherence of many OB associations (e.g. \citealt{QuintanaWright2021}) expected for a young group of stars that formed together, many studies have attempted to identify new OB associations from scratch. \textit{Gaia} data, combined with recent clustering techniques like the HDBSCAN algorithm \citep{HDBSCAN} or more sophisticated approaches such as the SigMA algorithm in \citet{Ratzenbock} and the code developed for the Villafranca project \citep{MaizApellaniz2019b}, allows the identification of new, kinematically-coherent and reliable OB associations (e.g. \citealt{QuintanaWright2021,Chemel2022,Quintana2023,Fleming2023,SaltovetsMcSwain2024}). 
\end{enumerate}
It is becoming increasingly clear that, aside from a few well-studied OB associations (e.g. Sco-Cen, Ori OB1, Vela OB2 and Cyg OB2), the historical lists of OB associations \citep{Humphreys1978,BlahaHumphreys1989} contain plenty of poorly-analysed groups, asterisms that need to be updated. 


\begin{figure}[h]
    \centering
    \includegraphics[scale =0.23]{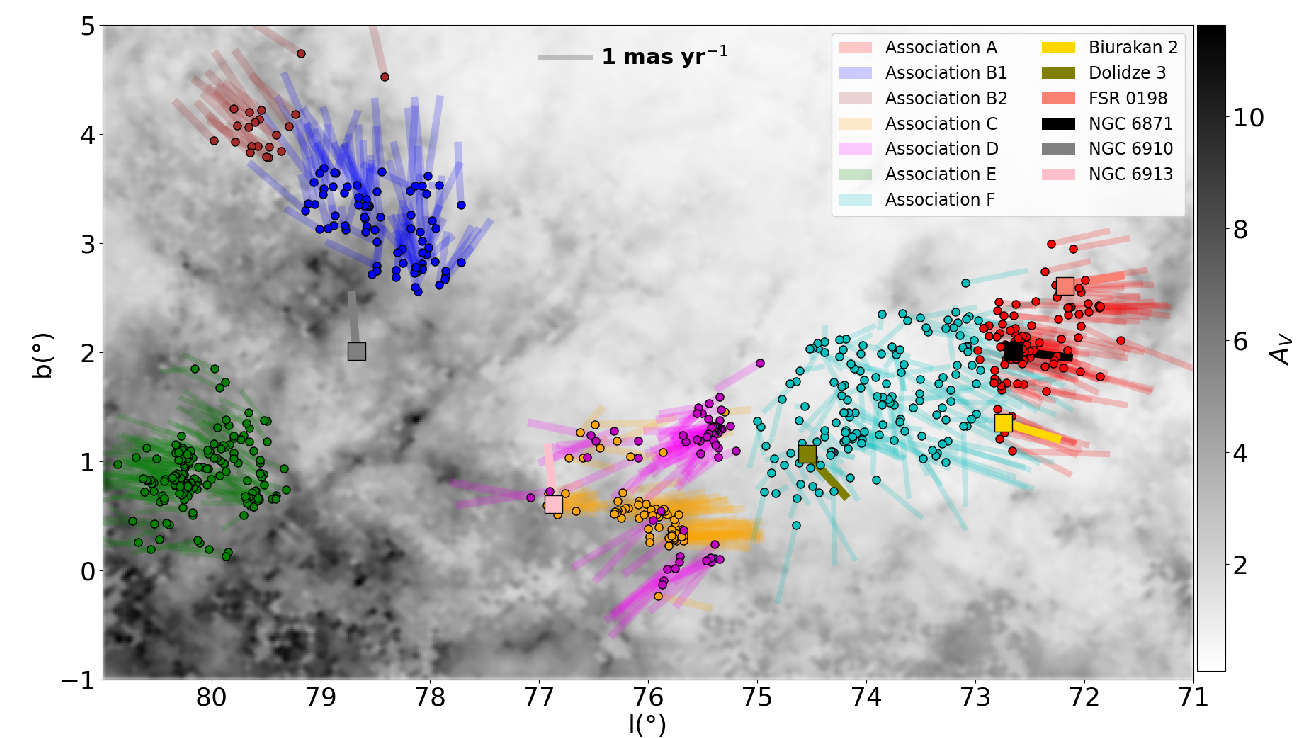}
    \caption{Galactic coordinates of the new Cygnus OB association members discovered in \citet{QuintanaWright2021} and their related OCs, with their relative PMs as vectors (observed PMs subtracted by the median PMs of the members). The background extinction map is from \citet{Bayestar2019}, while this figure has been taken from \citet{QuintanaWright2022}.}
    \label{OBAssocGaia1}
\end{figure}

Tremendous progress has thus been accomplished, both qualitatively and quantitatively. Nevertheless, I believe the history does not end here. Although \textit{Gaia} represents a giant leap forward compared with Hipparcos, allowing us to analyse OB associations at greater distances than ever, the precision on \textit{Gaia} parallaxes for studying OB stars starts to decrease around $\sim$3--4 kpc (e.g. \citealt{Quintana2023}). Thus, further developments are still halted by the available observations. This leads to one important question: what does the future entails, now that the past has been written? I will briefly discuss this in Section \ref{conclusions}. 

\section{From stellar to galactic scales}

The core of this review is dedicated to an exploration of OB associations from the scale of individual stars to the whole Milky Way. Extragalactic OB associations have been detected in the Magellanic Clouds (e.g. \citealt{HodgeWright1967,HodgeWright1977}), in the Andromeda Galaxy (e.g. \citealt{Hodge1981}) and other galaxies in the Local Group (e.g. \citealt{Hodge2002}). Nonetheless, their analysis depends upon completely different techniques, and as I aim this review to remain centered on observations from \textit{Gaia} and Galactic spectroscopic surveys, I will not explore beyond the limits of our Galaxy. I however invite interested readers to look through \citet{Gouliermis2018} for a review on the subject of extragalactic OB associations. 

Here, I will examine OB associations across various scales, not only emphasizing what we can learn from them, but also how they can be exploited to better comprehend many fields of astronomy.

\subsection{Massive stars in OB associations}
\subsubsection{Identifying massive stars in OB associations}

The initial step to establish a solid membership for OB associations is to reliably identify their individual stars. It is not rare to select low-mass stars to that end, especially in recent years, but such objects require further youth indicators (such as the presence of lithium from spectroscopy, e.g. \citealt{Armstrong2020}). This is why I will rather focus on the high-mass members, highlighting the progresses enabled by \textit{Gaia} and recent spectroscopic surveys. 

The historical approach consists of finding OB stars as per their \textit{classical} definition - that is, a massive star (initial mass greater than 9-12 M$_{\odot}$, e.g. \citealt{Poel,Langer}). Given their tremendous brightness and their scarcity, together with the original definition of OB associations, it is not surprising that they have been used to delineate OB associations (e.g. \citealt{Ruprecht1966,Humphreys1978}). This includes OB main-sequence stars as well as BA supergiants \citep{GarmanyStencel1992,Garmany1994}.

More recently, the advent of spectroscopic surveys such as the Galactic
O-Star Spectroscopic Survey (GOSSS) \citep{MaizApellaniz2011,Sota2011,Sota2014} has allowed us to characterize massive stars with unprecedented accuracy. These studies can be considered as an initial step towards an update on the mapping of OB associations, as their overdensities along the Galactic plane hints at their presence (e.g. in the Cygnus and Carina regions)

\begin{figure} [h]
    \centering
    \includegraphics[scale =0.35]{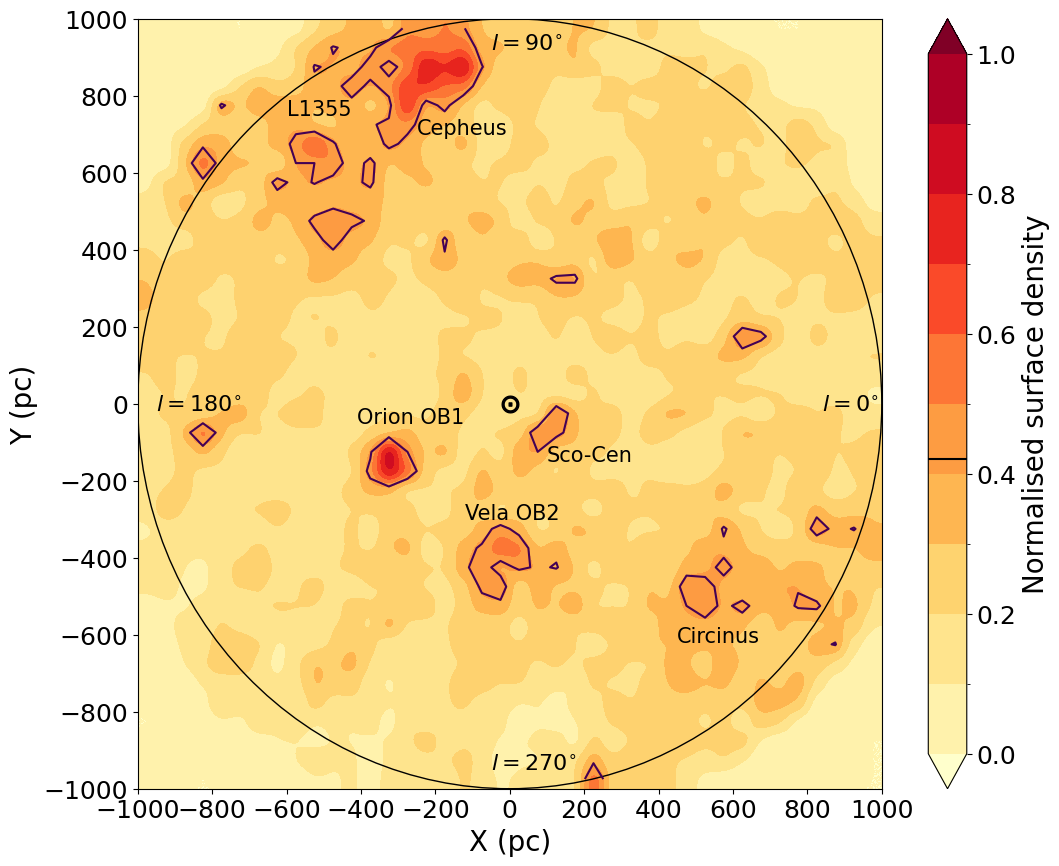}
    \caption{Normalised surface density of the SED-fitted OB stars from \citet{Quintana2024} in Cartesian coordinates, centered on the position of the Sun where the 24,706 SED-fitted OB stars within 1 kpc are contained within the circle. The contours represent overdensities likely to be real, with the selected threshold indicated on the colour bar. This figure was taken from \citet{Quintana2024}.}
    \label{OBmap}

\end{figure}

With the onset of \textit{Gaia} data, subsequent papers have henceforth incorporated \textit{Gaia} astrometry to perform an accurate 3D visualization of massive stars, for which there are several noteworthy examples. \citet{RateCrowther2020} updated the census of Galactic Wolf-Rayet (WR) stars with \textit{Gaia} DR2 data and, notwithstanding the greater isolation of WR stars compared with O-type stars, related their position to clusters and associations. Updating the Alma catalogue of luminous stars (ALS I, \citealt{Reed2003}) with \textit{Gaia} DR2, \citet{PantaleoniGonzalez2021} produced a robust 3D map of massive stars in the solar neighborhood, wherein are unveiled structures spanned by OB stars such as Vela OB1 and the Cepheus Spur. \citet{Zari2021} combined \textit{Gaia} EDR3 with 2MASS to map out luminous OBA stars in the Milky Way, identifying overdensities within the Cygnus, Cassiopeia and Carina regions amongst others. \citet{Quintana2024} applied an SED fitter (originally described in \citealt{QuintanaWright2021} and \citealt{Quintana2023}), relying on \textit{Gaia} DR3 photometry \& astrometry along with other photometry surveys, to identify and characterize $\sim$25,000 O- and B-type stars with $T_{\rm eff} >$ 10,000 K within 1 kpc from the Sun, whose normalised surface density in Cartesian coordinates is plotted in Figure \ref{OBmap}. Significant overdensities correspond to well-studied nearby OB associations such as Sco-Cen, Orion and Vela OB2, with the most striking structure, Cepheus, hinting at the presence of the Cepheus spur \citep{PantaleoniGonzalez2021}.

A clustering analysis on these stars is required to obtain a list of reliable OB associations (as in e.g. \citealt{MaizApellaniz2020,QuintanaWright2021}), yet these works have been crucial in efficiently locating probable sites of OB associations, from which a direct follow-up is easily attainable. 

\subsubsection{Characterizing OB associations through their massive members}

Once identified, individual members can help inform the properties about their parent association. Cyg OB2 is probably the most prominent example. \citet{Negueruela2008} observed obscured O-type stars together with luminous supergiants in Cyg OB2, from which they determined an age of $\sim$2.5 Myr for the association. \citet{Wright2015} produced a census of 169 OB stars in Cyg OB2, from which they estimated the age of the association (1--7 Myr), its mass function (with a power-law slope of $\Gamma = 1.3$) and a total mass of $16,500^{+3800}_{-2800}$ M$_{\odot}$. Likewise, \citet{Berlanas2018} conducted a spectroscopic study of Cyg OB2, discovering new massive members and revealing the high extinction in the area. It is notably through the study of stars such as J20395358+422250, a highly reddened blue supergiant undergoing huge mass-loss (for its spectral type, BIa), that bears out Cyg OB2 as an extreme environment \citep{Herrero2022}.

Similar work, where the stellar content of OB associations has been exploited to constrain their properties, has been carried out for Sco-Cen \citep{PreibischMamajek2008,Pecaut2012,Damiani2019}, Car OB1 \citep{Preibisch2011,Gvaramadze2020}, and Canis Major OB1 \citep{Gregorio,SantosSilva2018}, amongst others.

\subsection{Stellar multiplicity in OB associations}
\label{multiplicity}
If we extend our view slightly beyond individual stars, we can envision binary (and higher-order) stellar systems. Given that the multiplicity fraction increases as a function of the mass of the primary star \citep{DucheneKraus2013}, massive stars are more likely to belong to multiple systems than solar and low-mass stars (see e.g. \citealt{Duquennoy1991,Lada2006}). This signifies that OB associations provides suitable targets to study stellar multiplicity, learn how such systems emerge and evolve, and constrain their formation (e.g. \citealt{Barba2020}).

\subsubsection{Identifying multiple systems in OB associations}
\label{identifymultiple}

Studies of binary (and higher-order) populations of well-known OB associations have been carried out for decades, with Sco-Cen, the nearest OB association, being a favoured target (e.g. \citealt{Levato1987,Kraus2005}). From a sample of intermediate-mass primaries within Sco OB2, \citet{Kouwenhoven2007} found a binary fraction higher than 70 \% as well as a mass ratio distribution that favoured a power-law with an index of $\gamma = -0.4$. \citet{Rizzuto2013} conducted a interferometric survey of 58 B-type stars from the association and identified 24 previously-undetected companions, from which they could derive a mass ratio distribution with a power-law index of $\gamma = -0.46$ (thus close to the value from \citealt{Kouwenhoven2007}) alongside a companion frequency of $f = 1.35 \pm 0.25$. In a similar manner, \citet{Gratton2023} discovered 200 companions amongst 181 B-type stars in Sco-Cen, surmising that $\sim$75 \% of B-type stars are part of multiple systems and that massive (early B) stars tend to be in compact (i.e. with small angular separations) systems with massive secondaries. A similar conclusion was reached in \citet{Pauwels2023} for Sco OB1, since they found that the most massive association members were typically characterized by smaller angular separations, and estimated a multiplicity fraction of $0.89 \pm 0.07$ for the 20 OB stars in this association.

With its important massive star population, Cyg OB2 constitutes another preferred site to analyse stellar multiplicity. Cyg OB2 indeed contains prominent massive binaries, some of which have been individually studied such as Cyg OB2 \#22, a O3If+O6V binary system (e.g. \citealt{Walborn2002}), as well as Cyg OB2 \#8A, a O6I+O5.5III binary system (see e.g. \citealt{DeBecker2004}). \citet{Kiminki2007} realized a radial velocity survey of 146 OB stars in this association, notably uncovering several tens of spectroscopic binaries. \citet{Kiminki2012} determined an upper limit for the binary fraction of $90 \pm 10$ \% in Cyg OB2. A more recent census was carried out in \citet{CaballeroNieves2020}, with 47 \% of targets amongst a list of 74 early B and O-type stars being resolved companions.

\begin{figure}
    \centering
    \includegraphics[scale=0.4]{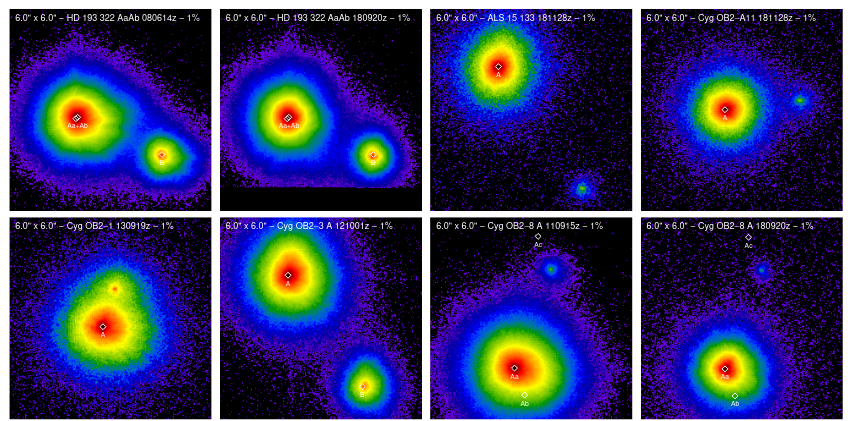}
    \caption{AstraLux fields for several massive stellar systems of the MONOS project, taken from \citet{MaizApellaniz2019} and reproduced with permission from EDP Sciences.} 
    \label{AstraLuxMONOS}
\end{figure}

Numerous techniques have been developed to identify multiple systems in OB associations, but a noticeable one is AstraLux, an imaging camera on the 2.2 m telescope of the Calar Alto observatory. In \citet{MaizApellaniz2010}, it was combined with High Resolution Channel (HRC) from the Hubble Space Telescope to obtain high-resolution imaging of several massive stars in OB associations, such as Cyg OB2 and Orion, where bound pairs of massive stars were found. Similarly, with the aim of better understanding the formation of massive stars, \citet{Peter2012} detected intermediate and massive binaries in Cep OB2 and Cep OB3 using AstraLux: besides favouring the competitive accretion model of star formation (e.g. \citealt{Bonnell2001}), they concluded that the environment does not influence the higher multiplicity fraction of massive stars. AstraLux was also exploited for the MONOS project: \citet{MaizApellaniz2019} led a spectroscopic analysis of $\sim$100 O-type stars from the GOSSS catalogue, that they complemented with visual imaging from the AstraLux camera (Figure \ref{AstraLuxMONOS}).

\subsubsection{Stellar multiplicity in low-density environments}

Once again, it is important to not solely envision OB associations as stellar groups delineated by their most massive members, but also as systems with their unique characteristics. One of their key features is their low density: therefore, one could expect that stellar multiplicity within OB associations exhibit different properties compared with dense star clusters. This idea has been suggested by \citet{Zinnecker2003}: dynamical interactions, combined with protostellar collisions, should favour the apparition of multiple systems. Specifically, wide binaries should be more frequently observed in OB associations, because they are less prone to disruptions when the stellar density is lower \citep{KroupaBurkert2001}. As stated in Section \ref{identifymultiple}, the presence of wide binaries in Cyg OB2 \citep{CaballeroNieves2020}, compared with dense clusters such as the ONC \citep{Scally1999}, reinforces this claim. Moreover, secular decays and close encounters should decrease the multiplicity fraction of stellar groups over time, and the fact that this quantity is typically high for massive stars in OB associations (e.g. \citealt{Kiminki2012,Pauwels2023}) further support it. Nevertheless, there is still a lack of statistically significant samples of massive binaries within OB associations across all angular separations, making the picture unresolved: I refer to the discussion from \citet{Wright2023} for additional details.

\subsection{Subgroups and open clusters}
\label{subgroupsoc}
At the parsec scale, there are two relevant, albeit supposedly distinct, entities within OB associations: their constituent, compact subgroups, as well as open clusters (hereafter, OCs). They have been of increasing importance in recent years, with \textit{Gaia} data widely applied to study their distribution, structure and age.

\subsubsection{Open clusters}
\label{oc}

When discussing the two main scenarios for the origins of OB associations in Section \ref{intro}, I contrasted their distinctive properties (gravitationally unbound and low-density) with those of OCs (bound and compact). In truth, OB associations often encompass OCs \citep{Wright2020}. Furthermore, although OCs can live much longer than OB associations, OCs serve equally as tracers of star formation in the Milky Way \citep{LadaLada2003,CantatGaudin2022,CantatGaudinCasamiquela2024}. 

\begin{figure} [h]
    \centering
    \includegraphics[scale =0.25]{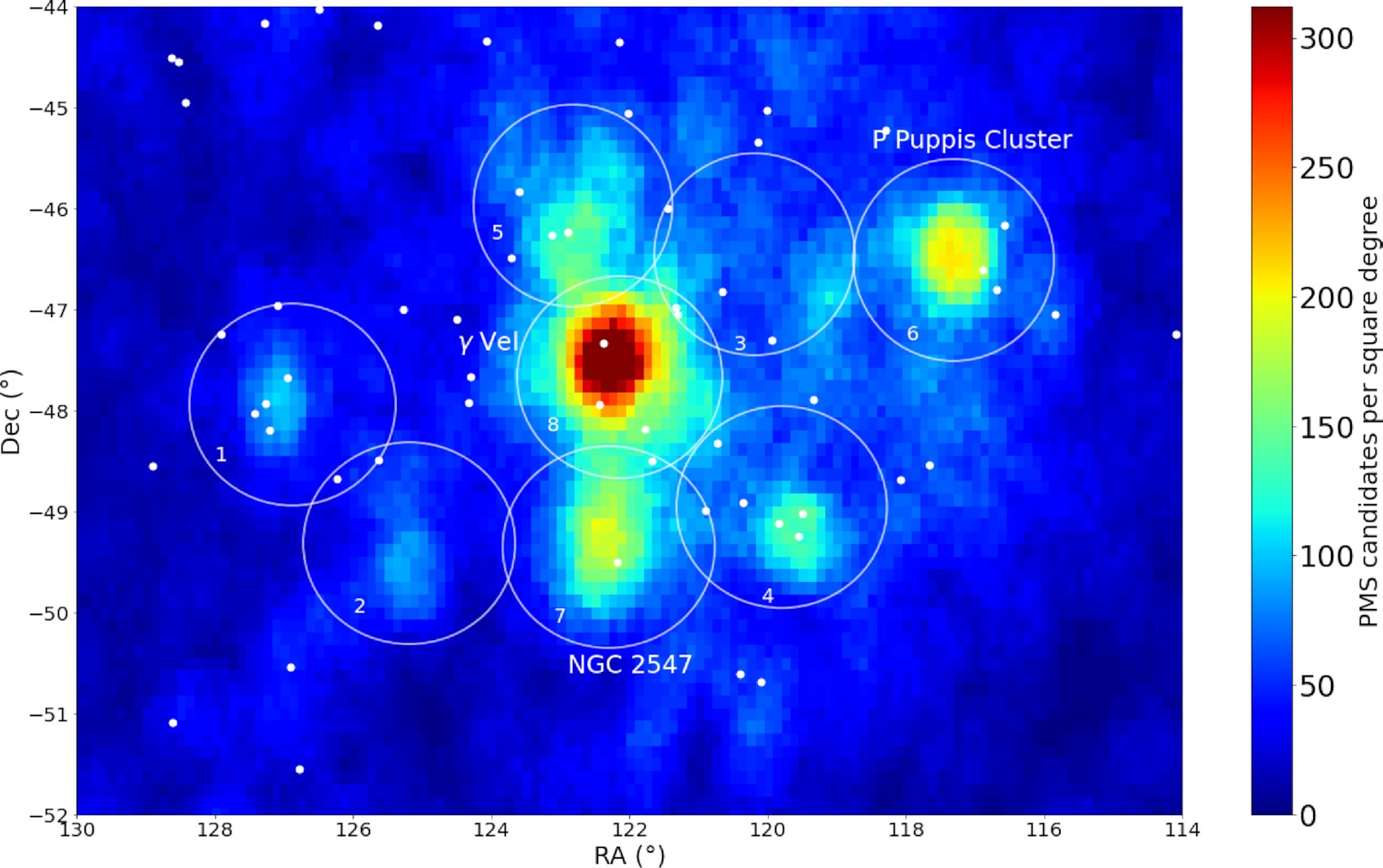}
    \caption{Equatorial coordinates of Vela OB2, its constituent OCs (with $\gamma$ Vel particularly standing out) as well as the OB members of Vela OB2 from \citet{HipparcosCensus1999} shown as white dots, plotted on top of a density map of the candidate PMS stars from \citet{Armstrong2022}.}
    \label{VelaOB2OCs}
\end{figure}

The original classification of OB associations frequently includes related OCs (see \citealt{Humphreys1978} and \citealt{Wright2020}). Noticeable examples of OCs are $\gamma$ Velorum in Vela OB2 (e.g. \citealt{Jeffries2014,Armstrong2022}, also see Figure \ref{VelaOB2OCs}), Stock 8 in Aur OB2 (e.g. \citealt{NegueruelaMarco2003,MarcoNegueruela2016}) and NGC 663 in Cas OB8 (e.g. \citealt{FabregatCapilla2005,Yu2015}).

OCs can thus be utilized to learn about the OB association they belong to. For example, a photometric study of Stock 8 and NGC 1893 in Aur OB2 revealed that, contrary to what was formerly thought, these OCs are not located at the same distance, encouraging a revision of the classical definition of Aur OB2 \citep{MarcoNegueruela2016}. \citet{Quintana2023} went on to identify new OB associations in Auriga, exploiting related OCs to constrain their properties.

\subsubsection{Subgroups}
\label{sub}

The existence of subgroups in OB associations has been established for decades (e.g. \citealt{Blaauw1946,Blaauw1964,GarmanyStencel1992}), but the rate at which they have been discovered and examined has truly risen since \textit{Gaia}.

Perhaps the most famous examples of such subgroups are in Sco-Cen: being the closest OB association \citep{HipparcosCensus1999}, its subgroups stand out clearly over the plane-of-the-sky. Initially discovered by \citet{Blaauw1946}, Upper Scorpius (US), Upper Centaurus-Lupus (UCL) and Lower Centaurus-Crux (LCC) are characterized by distinct and different median ages (11 to 17 Myr, \citealt{PecautMamajek2016}) that reveal the star formation history of their parent association. In light of \textit{Gaia} kinematics, these subgroups unveil a deeper substructure: as evoked in \citet{WrightMamajek2018}, the historical division of Sco-Cen has now been superseded, thus subsequent papers such as \citet{Squicciarini2021} delved deeper into the kinematics of Sco-Cen. Conspicuous developments are the 7 groups found in Upper Scorpius and Ophiuchius in \citet{MiretRoig2022}, as well as the 37 groups in Sco-Cen from \citet{Ratzenbock}. In both studies, the age differences between the subgroups served as the starting point to analyse the propagation of star formation inside the regions and the mechanisms responsible for the creation of these patterns. 

Another prominent case of Ori OB1 which \citet{Blaauw1964} originally divided between four subgroups, whose stellar content and age were further studied in \citet{Brown1994}. Just like Sco-Cen, the arrival of \textit{Gaia} opened an updated view of the association, with the detection of a rich young stellar population alongside the presence of a plethora of subgroups forming an age gradient and parts of larger structures \citep{Kounkel2018,Zari2019,Sanchez2024}.

\begin{figure}[h]
    \centering
    \includegraphics[scale =0.22]{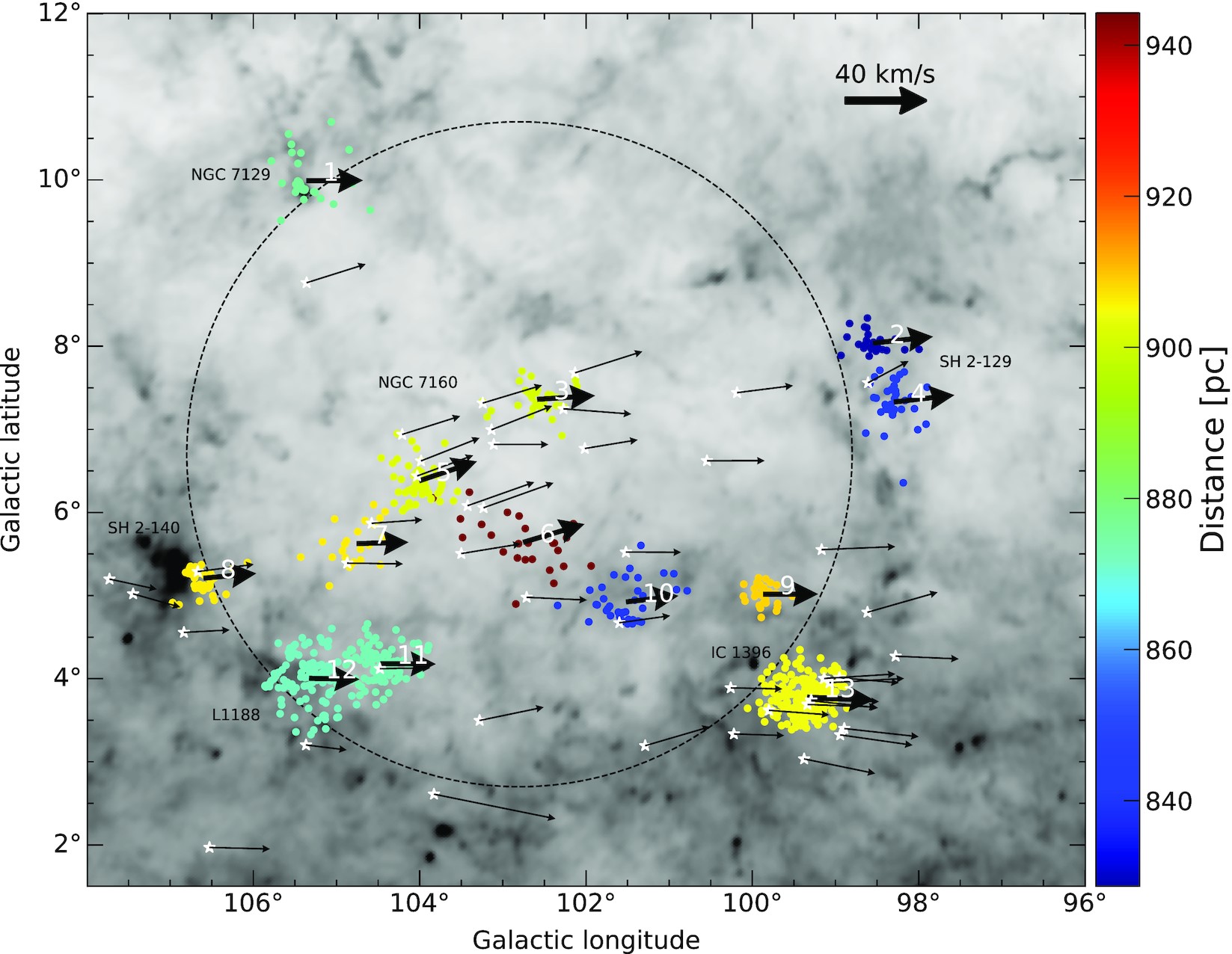}
    \caption{Members of the subgroups of Cep OB2 identified by HDBSCAN in \citet{Szilagyi2023}, colour-coded by distance, whose vectors correspond to the tangential velocities, and whose massive stars ($>$ 10 $M_{\odot}$ here) have been displayed as white dots.}
    \label{CepOB2subgroups}
\end{figure}

Recently, subgroups have also been identified, and their age distribution studied, in other OB associations, noticeably in Vela OB2 (e.g. \citealt{Cantat2019}), Sco OB1 (e.g. \citealt{Yalyalieva2020}), Mon OB1 (see e.g. \citealt{Lim2022}) and Cep OB2 (e.g. \citealt{Szilagyi2023}, see Figure \ref{CepOB2subgroups}). Likewise, the Villafranca project combined \textit{Gaia} data with the spectroscopic information from the GOSSS survey \citep{Sota2011,Sota2014} to identify OB groups, several of them catalogued as part of wider OB associations such as Ori OB1 and Cyg OB2 \citep{MaizApellaniz2020,MaizApellaniz2022,Villafranca}.

\subsubsection{Distinction}
\label{delim}

Distinguishing between OCs and subgroups is, in principle, straightforward: a core property of OCs is that they are gravitationally bound, whereas the bulk of subgroups are gravitationally unbound. In practice, however, these entities can be related: the Villafranca project favoured the label `OB group' even though they extended over a cluster scale, while \citet{Yalyalieva2020} and \citet{Szilagyi2023} connected their identified subgroups to surrounding OCs. Likewise, despite preferring the term `cluster', \citet{Ratzenbock} specified that all their groups identified in Sco-Cen should be gravitationally unbound, akin to the definition of a subgroup, and a similar approach was followed in \citet{Posch2024}. Likewise, \citet{Swiggum2024}, who combined the star clusters from \citet{HuntReffert2023} (regardless of their state of boundness) with the Young Local Associations (YLAs, defined as coeval and low-mass groups of stars closer than 200 pc) from \citet{Gagne2018}, referred to everything as `clusters'.

Often the studies of OCs and subgroups within OB associations break down into a kinematic analysis where the boundness is not considered. \citet{HuntReffert2024} calculated the Jacobi radius and effectively separated their list of clusters into 5647 bound OCs and 1309 unbound clusters (referred to as `moving groups', hereafter MGs). Could these MGs be the subgroups of OB associations?

\begin{figure}[h]
    \centering
    \includegraphics[scale = 0.35]{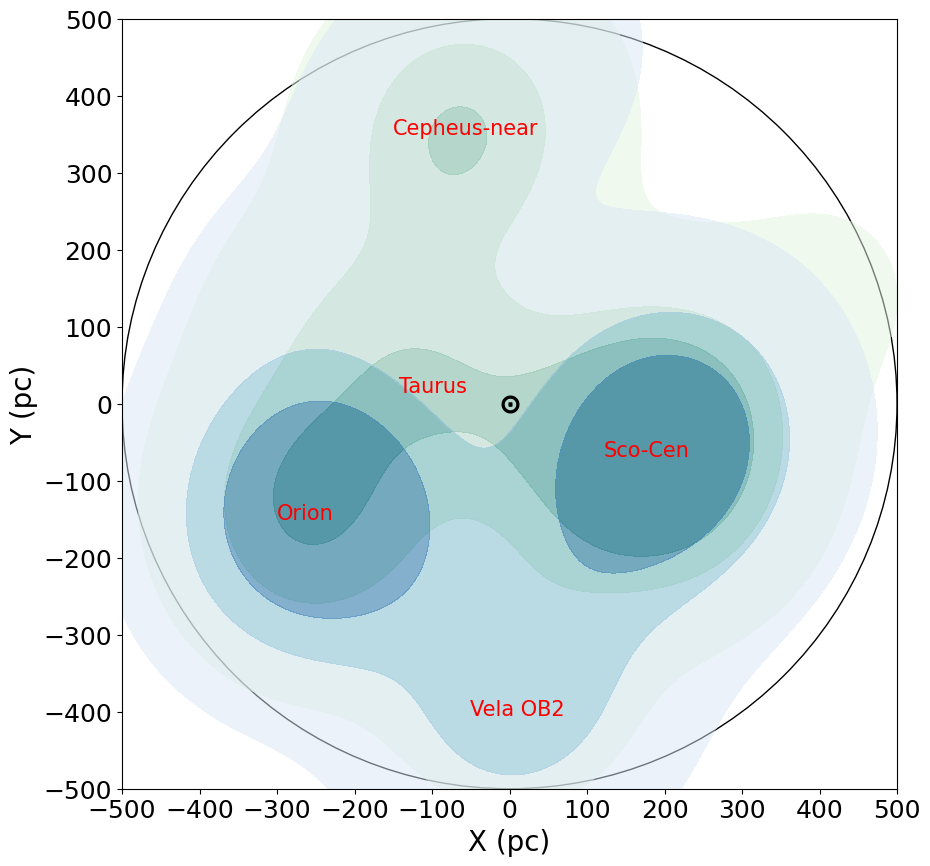}
    \caption{Kernel density estimates with four level contours of the 38 OCs and 23 MGs of high-quality and young age ($<$ 10 Myr) within 500 pc, from \citet{HuntReffert2024}. The positions of the known OB associations \citep{Wright2020} and noticeable star-forming regions \citep{Zucker2020} have been indicated.}
    \label{HR24Map}
\end{figure}

To investigate this, I have taken the 38 OCs and 23 MGs younger than 10 Myr, with $\sqrt{X^2+Y^2} <$ 500 pc and of high-quality (defined in \citealt{HuntReffert2023} as clusters with a CMD class greater than 0.5 and an astrometric SNR greater than 5) from \citet{HuntReffert2024}, and shown them with the major known OB associations (from \citealt{Wright2020}) and star-forming regions (from \citealt{Zucker2020}) in Figure \ref{HR24Map}. Highlighted here is the relation between Sco-Cen and Taurus, consistent with \citet{Swiggum2024} who found that they belong to a same family, and Alpha Persei, with its most compact state $\sim$20 Myr ago. Overdensities of OCs and MGs overlap within Sco-Cen, Orion and Vela OB2, suggesting that they are composed of both bound OCs and unbound subgroups. By contrast, Taurus and Cepheus-near are only coincident with overdensities of MGs. 

The term `moving group' itself has been debated: the presence of an overdensity of MGs in Taurus, a low-mass star-forming region, could imply that they correspond to YLAs there, and would only be unbound subgroups of OB associations if they are found alongside overdensities of OCs. However, it is hard to draw a definitive conclusion from this picture, as it is limited to 500 pc from the Sun. 

\subsection{Feedback, H{\sc ii} regions and interactions with the ISM}
\label{feedback}

Sources of stellar feedback comprise protostellar outflows, powerful stellar winds, photoionization, radiation pressure and supernova explosions, whose origin can be found in OB stars (e.g. \citealt{Krumholz2014,Dale2015}). All these phenomena exert a significant influence on their surrounding environment: they contribute to the emergence of H{\sc ii} regions (e.g. \citealt{Anderson2009}) and halt star formation by dispersing the surrounding molecular gas \citep{DobbsPringle2013,Krumholz2019}. Stellar feedback is the primary cause of destruction of GMCs, as this occurs over a scale of a few Myrs, long before the GMC can convert its gas into stars \citep{Chevance2020}.

OB associations live up to a few tens of Myrs and will therefore witness several core-collapse supernova explosions from their most massive members during their existence. The joint effect of each detonation will blow away the surrounding interstellar material and generate cavities of large size ($>$ 100 pc), very high temperature ($>$ 10$^6$ K) and extremely low density ($<$ 0.01 cm$^{-3}$) \citep{Higdon2005,Higdon2013,Drozdov2022}. 

There are many observations of superbubbles encompassing OB associations, such as in Cyg OB2 \citep{Cash1980}, Ori OB1 \citep{ReynoldsOgden1979} and Cep OB2 \citep{Kun1987}. More recently, the HaloSat instrument was able to detect high temperature gas and energy injection around Cygnus \citep{Bluem2020} and Orion \citep{Fuller2023}. Subgroups from Cep OB2 were also related to the expansion of the Cepheus superbubble in \citet{Szilagyi2023}, as the younger groups are located at the periphery of the association, thus possibly forming from collisions between their parent molecular cloud and the expanding superbubble.

\begin{figure}[h]
    \centering
    \includegraphics[scale= 0.35]{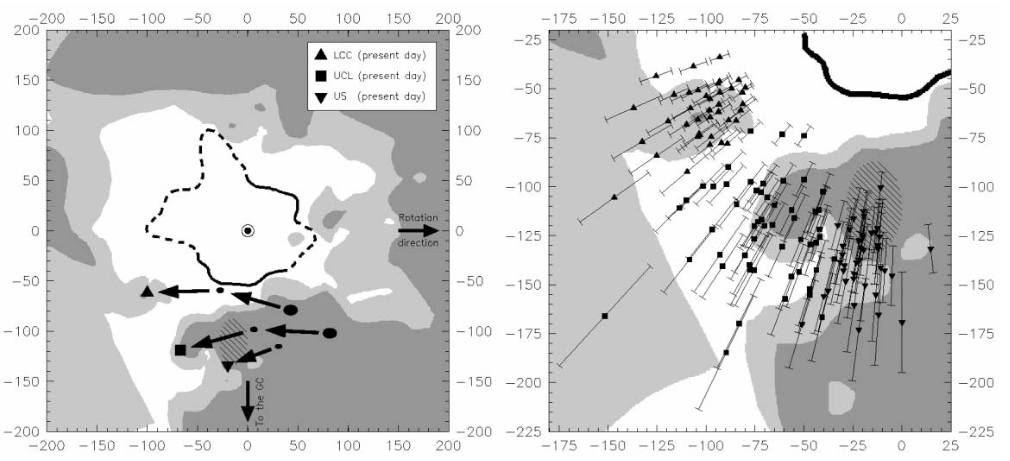}
    \caption{Left panel: The Local Bubble and local cavity as displayed on the plane of the Galactic equator, alongside the past and present position of the subgroups of Sco-Cen using Hipparcos astrometric data. The contours delimit the present size of the LB from X-ray data \citep{Snowden1998}, while the fill contours show the Na I distribution \citep{Sfeir1999}. Right panel: Zoomed view of the left panel, where the OB stars from the subgroups with a reliable position have been shown. This figure was taken from \citet{MaizApellaniz2001} (DOI: 10.1086/324016). ©AAS. Reproduced with permission.}
    \label{LocalBubble}
\end{figure}

The Local Bubble (LB) itself may have originated from an OB association. Using astrometric data from Hipparcos, \citet{MaizApellaniz2001} showed that Sco-Cen was located at the Sun's position 5--7 Myr ago, and estimated through evolutionary analysis that Sco-Cen experienced up to 20 supernova explosions over the past 10-12 Myrs, during which the resulting feedback could have produced the LB. This is illustrated in Figure \ref{LocalBubble}. \citet{Zucker2022} have validated this scenario using \textit{Gaia} data, though placing the beginning of these supernova explosions $\sim$14 Myr ago instead.

Simulations also support the view of superbubbles generated by OB associations: in \citet{Drozdov2022}, twenty supernova remnants (SNRs) within the galactic disk were connected to OB associations. They found that the morphology of the SNRs notably depends upon the radius of the clusters as well as the mass function of its members.

On the other hand, supernova explosions from OB association members inject new chemical elements into the ISM, which are used to engender subsequent generations of stars. Whether OB associations constitute sources of self-enrichment, enhancing surrounding gas through their stellar winds (e.g. \citealt{Decressin2007,Nowak2022}), abides as a debate. For instance, \citet{CunhaLambert1994} argued that the young subgroups in Ori OB1 were more abundant in oxygen, which was challenged by \citet{SimonDiaz2010} who found a higher degree of homogeneity in the Si and O abundances within these subgroups, therefore opposing the self-enrichment scenario.

\subsection{Galactic structure}
\label{galacticstructure}

\begin{figure} [h]
    \centering
    \includegraphics[scale = 0.9]{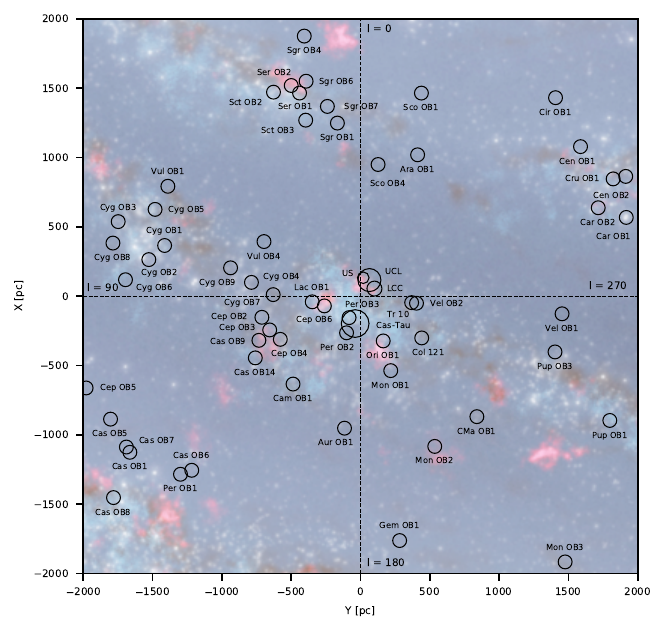}
    \caption{The classical lists of OB associations displayed in Cartesian coordinates within 2 kpc (using their \textit{Gaia} DR2 distances), with the background representing an artistic view of the Milky Way spiral arms; respectively, from bottom to top, the Perseus, Local and Sagittarius Arm (Credit: NASA/JPL-Caltech/R. Hurt). Figure courtesy of Nick Wright, adapted from Figure 19 in \citet{Wright2020}.}
    \label{OBAssocCartesian}
\end{figure}

The final step consists of visualising OB associations across the scale of the entire Milky Way. Because spiral arms are important sites of star formation (e.g. \citealt{Elmegreen2011}), and because OB stars remain close to their birth environment due to their brief existence (e.g. \citealt{SparkeGallagher2000}), they represent valuable tracers of Galactic structure (e.g. \citealt{Russeil2003,Chen2019,PantaleoniGonzalez2021,Zari2021}). Therefore, unsurprisingly, many OB associations sit within the spiral arms, as illustrated in Figure \ref{OBAssocCartesian}.

It is however crucial to point out that, since classical lists of OB associations need to be revised, so is their distance: to give an example, the historical Cepheus OB associations are situated closer than the overdensities of OB stars in the Cepheus region from Figure \ref{OBmap}. Additionally, \citet{NegueruelaMarco2003} exploited various OB associations as tracers of the Galactic spiral arms, and whilst some of them were good tracers (e.g. Cam OB3 for the Outer Arm), this study served as an early example that the definition and extent of Aur OB2 needed to be revisited. 

This is where \textit{Gaia} data is powerful, as it enables us to map out overdensities of OB stars with a greater accuracy than ever before, confirming or confuting previously outlined structures, as well as discovering new ones. For instance, the 3D mapping of young stars with \textit{Gaia} DR2 in \citet{Zari2018} failed to pinpoint the OB groups identified in \citet{BouyAlves2015} with Hipparcos data, while \citet{Zari2021} did recover similar structures to \citet{Wright2020} (at least nearby) by mapping out the overdensities of OBA stars using \textit{Gaia} EDR3 and 2MASS data. \citet{Zari2021} also found that the structures unveiled by these overdensities did not exactly correspond to the positions of the spiral arms, but this could ascribed to the fact that this catalogue includes stars that are sufficiently old to have moved away from their birth environment. On the other hand, mapping out the 3D distribution of massive (OB) stars in the Milky Way, \citet{PantaleoniGonzalez2021} spotted a kinematically distinct structure referred to as the Cepheus spur, extending from Orion-Cygnus towards the Perseus spiral arm. 

\begin{figure}
    \centering
    \includegraphics[scale = 0.21]{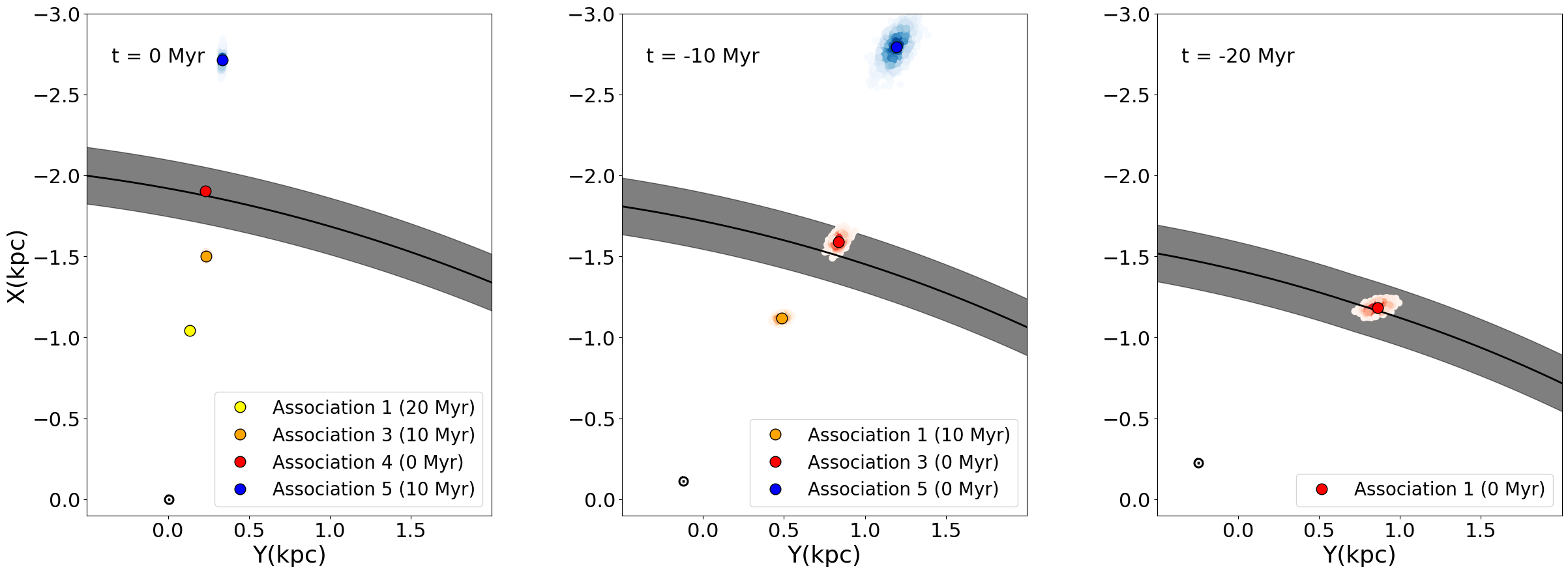}
    \caption{Snapshots of the 5 new OB associations in the Auriga region (in Cartesian coordinates), from the present time to 20 Myr ago, where the age of each OB association at these times is indicated in the legend. Their location has been compared with the Perseus spiral arm (whose position and thickness have been taken from \citet{Reid2019}), using the equation from \citet{Dias2019} to trace back its motion over time. The age of the OB associations have been constrained with independent techniques (positions of their members in the HR diagrams and age of their related OCs). This figure has been adapted from \citet{Quintana2023}.} 
    \label{AurPerseus}
\end{figure}

These recent studies suggest that OB associations are promising sites to disentangle Galactic structure, but I argue that we can go further for future developments of the field. OB associations typically live for a few tens of Myrs (e.g. \citealt{Blaauw1964}): this corresponds to the timescale over which we can perform an accurate dynamical traceback with current \textit{Gaia} data (e.g. \citealt{MiretRoig2022}). \citet{Quintana2023} identified an age gradient spanned by the new OB associations discovered in the Auriga region, that coincide with the motion of the Perseus spiral arm over the last $\sim$20 Myrs, as illustrated in Figure \ref{AurPerseus}.

OB associations could therefore serve as tracers of the past motions of the Galactic spiral arms, over many tens of Myrs. Consequently, just like OCs \citep{CastroGinard2021}, a 3D kinematic traceback of a reliable, all-sky list of OB associations could help to determine the rotation pattern speed of the spiral arms. This would contribute towards the fundamental question of the origin of the spiral arms, i.e. whether they are long-lived entities that rotate like rigid bodies at a constant angular speed, independently from the stars (\textit{density wave theory}, \citealt{LinShu1964}); or if they correspond to transient structures growing and fading away with time (e.g. \citealt{Quillen2011,Wada}), following a model where each spiral arm is characterized by their own, distinct rotation pattern speed. 

\section{Summary and future perspectives}
\label{conclusions}

The field of OB associations is rapidly evolving, impulsed by the advent of precise and large-scale astrometric, photometric and spectroscopic surveys, among which \textit{Gaia} particularly stands out. Even though these recent studies only represent a fraction of the thousands of publications exploiting \textit{Gaia} data, there have been nonetheless pivotal to improve our understanding of star formation, stellar multiplicity, feedback and Galactic structure.  

A plethora of open questions remain, such as whether multiple systems exhibit different properties compared with dense star clusters and how to properly distinguish between compact subgroups and OCs within OB associations. It goes without saying that future facilities will help us to solve these mysteries. 

Forthcoming advances will come for the subsequent \textit{Gaia} data releases: \textit{Gaia} DR4 is currently scheduled for mid-2026, with PMs expected to be $\sim$2.8 times more precise than those of \textit{Gaia} DR3 \citep{Brown2019}. Awaited for late 2030 at the earliest, \textit{Gaia} DR5 will be the final data release of this mission, from which we can anticipate more improvements in the kinematics of OB associations. 

Further to this, one of the present limitations of the field is the lack of RVs for OB stars: not only are the RVs from \textit{Gaia} DR3 limited to the late B-type stars (up to 14,500 K), but their precision only reach $\sim$10 km s$^{-1}$ at this range \citep{Katz2023}. Upcoming large-scale spectroscopic surveys are predicted to solve this issue. 4MOST is set to obtain spectra for $\sim$10,000 OB stars in the southern hemisphere \citep{4MOST} whilst its equivalent of the northern hemisphere, WEAVE, is expected to acquire spectra of a similar number of OB stars. In particular, the SCIP survey will observe $\sim$400,000 OBA stars and YSOs across the Galactic plane, with high-resolution spectroscopic monitoring of the Cygnus region and Galactic anticentre \citep{WEAVE}. In addition, the BOSS survey from SDSS-V should measure spectra of $\sim$100,000 OB stars by 2028 \citep{Kounkel2023}, and there will be soon the updated versions of the ALS and GOSSS catalogues, providing additional spectroscopic information \citep{Villafranca}.

The area of OB associations will undoubtedly benefit from the data of these upcoming facilities. On the one hand, as a follow-up, they will be able to confirm the nature of OB(A) stars identified through astro-photometric techniques (e.g. \citealt{Zari2021,Quintana2024}), by providing accurate stellar parameters for these targets. On the other hand, their RVs will be an asset, both from a qualitative and a quantitative point of view: 4MOST and WEAVE should attain precision down to 1 km s$^{-1}$ for their OB stars \citep{4MOST,WEAVE}, enabling unprecedented 3D kinematical analysis on OB associations. 

Future photometric surveys are also promising. The LSST/Rubin Observatory will be $\sim$3 magnitudes deeper than \textit{Gaia} (and even $\sim$6 mag in co-added images), thereby providing more photometry as well as proper motions for fainter sources \citep{LSST}. In addition, the Roman Galactic Bulge Survey will be able to penetrate deep into the Galactic centre, possibly resolving distant OB associations (even though crowding will be an issue, see e.g. \citealt{Roman}).

The telescope that could further revolutionize the field of OB associations is GaiaNIR. As follow-up of the \textit{Gaia} mission, GaiaNIR is aimed at observing $\sim$12 billion stars in the near-infared, thereby increasing the \textit{Gaia} census by a factor of 6. On top of producing an updated astrometric catalogue, GaiaNIR is set to explore regions obscured by dust (notably the Galactic centre), unveiling stellar groups still embedded within their birth cluster, and thus the missing stage between star-forming regions and OB associations decoupled from their natal environment. GaiaNIR is yet to be approved by ESA, but if it occurs as scheduled, it will be launched around 2045, $\sim$1 century after OB associations were defined for the first time, timely for another revolution \citep{HobbsHog2018,Hobbs2024}.

\section*{Acknowledgements} 

I acknowledge the support from the Spanish Government Ministerio de Ciencia, Innovaci\'on y Universidades and Agencia Estatal de Investigación (MCIU/AEI/10.130 39/501 100 011 033/FEDER, UE) under grants PID2021-122397NB-C21/C22 and Severo Ochoa Programme 2020-2024 (CEX2019-000920-S). I also acknowledge the support from MCIU with funding from the European Union NextGenerationEU and Generalitat Valenciana in the call Programa de Planes Complementarios de I+D+i (PRTR 2022), project HIAMAS, reference ASFAE/2022/017.

I am grateful to Joseph Armstrong, Hervé Bouy, Bruce Elmegreen, Jesús Maíz Apellániz, Nick Wright, Máté Szilágyi for authorizing me to use their figures in this review. 

I am also thankful to Hervé Bouy, Anthony Brown, Emily Hunt, Núria Miret-Roig and Nick Wright for their careful reading of the manuscript and their comments that allowed me to improve the paper.


\bibliography{sample}
\bibliographystyle{ceab}


\end{document}